\documentclass[12pt]{iopart}

\usepackage{iopams}
\usepackage{bm}
\usepackage{graphicx} 
\usepackage{epstopdf}
\usepackage{cite}

\begin{document}

\title[Nodes and QHJ solution]{On the nodes of wave function and the quantum Hamilton-Jacobi solution}

\author{L A Poveda-Cuevas$^{1}$, F J Poveda-Cuevas$^{2}$}

\address{$^1$ Instituto de Matem\'{a}tica e Estat\'{i}stica, Universidade de S\~{a}o 
Paulo, C.P. 05508-090, S\~{a}o Paulo, SP, Brazil}
\address{$^2$ Instituto de F\'{i}sica, Universidad Nacional Aut\'{o}noma de M\'{e}xico,
C.P. 20-364, 01000 M\'{e}xico, Ciudad de M\'{e}xico}

\ead{lpovedac@ime.usp.br}
\ead{jacksonpc@fisica.unam.mx}

\vspace{10pt}
\begin{indented}
\item[]\today
\end{indented}

\begin{abstract}
We present the analytic solution for the stationary quantum Hamilton-Jacobi equation. Knowing the strong relation between the Riccati and quantum Hamilton-Jacobi equations, we develop a simple method to obtain the exact solution. Then, in order to prove the validity of the proposed method, we use two central potentials: the three-dimensional harmonic oscillator and Coulomb potential, both with bound-states. Finally, we compute the action-angle variables in a entirely quantum version for to achieve connect with the nodes of the wave function.
\end{abstract}


\vspace{2pc}

\submitto{\JPA}
%
%
%

\section{Introduction}

The main objective of quantum mechanics is to determine the energy spectrum. In this way, many formulations were developed along the last century to reach this aim (see e.g., \cite{Styer-ajp70} and the reference there in). One of the most interesting formulation is the Quantum Hamilton-Jacobi (QHJ), which is given by the nonlinear operator partial differential equations \cite{Roncadelli-prl99, Periwal-prl80, Castro-foundphys21}. In particular, Leacock and Padget point out the formalism of QHJ as useful tool to directly calculate the spectrum without solving the eigenfunctions problem \cite{Leacock-prl50,Leacock-prd28}. However, the power series method is used to seek a solution of linear differential equations, but the flexibility of the method is strongly reduced when to be applied to certain nonlinear differential equations. This is precisely the case of the QHJ equation, which is essentially a Complex Non-linear Riccati (CNLR) differential equation, where the quadratic term induces the non-linearity. The original proposal is to be used the Laurent's power series method in order to solve the QHJ equation; nevertheless these power series can not be applied directly by the following conditions: (i) the kind of singularities are not defined, (ii) the solution is not unique, because the superposition principle is not satisfied \cite{Teschl-book2012}. Further, another strategies to resolve nonlinear equations, such as the Parker\textendash Sochacki method involves a initial value condition, nonetheless the quantum mechanics has fundamentally boundary value problems \cite{Teschl-book2012}. 

Several methods have widely studied the case of non-linearity of CNLR \cite{Schuch-jphysconfseries504, Ahlbrandt-book1996, Reid-book1972}. Thus, the main focus of this work is to find the analytic solution for the stationary QHJ equation for different potentials through the CNLR equation. At the same time we are showing the \textit{equivalence} between the QHJ and Schr\"{o}dinger formalism. In particular, we illustrate the link up between of both equations with a few examples such as the harmonic oscillator and hydrogen atom. Usually in the literature the one-dimensional QHJ equation is widely studied by the interesting properties related with supersymmetry in quantum mechanics \cite{Ranjani-modphysletta19}; nevertheless the intent of this work is to show the richness of the problems in three-dimensions.

Recently the community has taken interest about the topological properties of the wave function and its relation with the nodes \cite{Zhoushen-prl117}. In fact, the debate about the nodes of wave function and its importance in one-dimensional problems can be found in the references \cite{Zettili-book2009, Moriconi-quant-ph.0702260}. In main strategy of this work, once we have the solution of the QHJ equation, our aim is to show how the nodes (zeros) and anti-nodes (maxima) of the wave function are connected via the angle-action variables. We show that the ground state wave-function has no nodes, even in the three-dimensional problem, and we obtain the nodes for exciting levels only analyzing the behavior of the solution obtained by the method presented in this article. Hence, we shown an explicit form of compute the nodes of the wave function, which correspond to roots where the momentum function is non-analytic on the complex plane. Also we show a brief analysis of how to get the anti-nodes of the wave function.

This article is organized as follow. In Section \ref{sec:QHJCNLR} we being by reviewing a few elements of current theory about the QHJ and CNLR equations and we introduce some notations. In the Sections \ref{sec:HarmonicOscillator} and \ref{sec:HydrogenAtom} we illustrate two classical examples, and we explicitly show some results about the momentum function and action-variable variables. Finally, conclusions are drawn in Section  \ref{sec:conclusion}.

\section{Quantum Hamilton-Jacobi and Ricatti equations}\label{sec:QHJCNLR}

\subsection{Overview of Quantum Hamilton-Jacobi}

It is clear from the classical mechanics \cite{Goldstein-book2002} that a system with the central potential can be separated by some suitable canonical coordinate transformation. This one reduces the original problem to a system of differential equations in terms of cyclic coordinates.
When the system is time-independent, we can separate explicity $E$ as constant of motion, therefore the energy is conserved and considered a cyclic coordinate \cite{Born-book1960}. Note that, the formulation requires the existence of conserved quantities or generalized cyclic coordinates. The existence of this cyclic coordinate indicates a conserved quantity and it does not explicitly appear in the Hamiltonian. This fact guarantees the separability of equation of motion. On the other hand, in the wavefunction formulation $\hat{\mathbf{q}}$ and $\hat{{\bf p}}$ have category of conjugated operators and satisfy the commutation relation, i.e., $\left[\hat{q}_{i},\hat{p}_{j}\right]=i\hbar\delta_{ij}$ \cite{Merzbacher-book1970}. From this point of view, the problem of dynamic of operators can be avoided using the formulation of QHJ. 

We consider the time-independent QHJ equation defined in \cite{Leacock-prl50, Leacock-prd28} as
\begin{equation}
-\frac{\mathrm{i}\hbar}{2{\rm m}}\nabla^{2}W+\frac{1}{2{\rm m}}\left(\bm{\nabla}W\right)\cdot\left(\bm{\nabla}W\right)=E-V\left(\mathbf{q}\right),\label{QHJW}
\end{equation}
where $\mathrm{m}$ is the mass of the particle and $E$ is the energy. We define the quantum characteristic function, 
\[
W\left(\mathbf{q},\bm{\kappa}\right) \equiv W,
\]
with $\mathbf{q}=\left(q_{1},q_{2},...,q_{f}\right)$ are the generalized coordinates and $\mathbf{\bm{\kappa}}=\left(\kappa_{1},\ldots,\kappa_{f-1},E\right)$ is a set of parameters, where $f$ represents the number of degrees of freedom. The separation of $W$ for a system with many degrees of freedom is typically given by a sum
\[
W=\sum_{i=1}^{f}W_{i}\left(q_i, \kappa_j \right)
\] for each coordinate $q_{i}$. In this manner, $\kappa_{j}$ is a separation constant for $j=1,\dots,f-1$. 

The central potential $V\left(\mathbf{q}\right)$ with bound states is considered in this work. We recall the QHJ equation (\ref{QHJW}) as
\begin{equation}
-\mathrm{i}\hbar\bm{\nabla}\cdot{\bf p}+{\bf p}\cdot{\bf p}=2{\rm m}\left(E-V\left(\mathbf{q}\right)\right).\label{eq:HJEpfunction}
\end{equation}
The new physical quantity ${\bf p}$ is a function implies a separability when
\begin{equation}
{\bf p}=\bm{\nabla}W,\label{eq:pfunc}
\end{equation}
where $\mathbf{p}=\left(p_{1},p_{2},...,p_{f}\right)$ are the associated momentum functions. The associated functions are not more operators, moreover they are totally analog to generalized canonical variables of classical mechanics. Hence, the dynamic of the state is completely described in terms of this set of functions. 

On the other hand, we can establish the following relation:
\begin{equation}
p_{i}=-\mathrm{i}\hbar\frac{1}{\psi}\frac{\partial\psi}{\partial q_{i}}.\label{eq:ppsifunc}
\end{equation}
The $\psi$ is the eigenfunction of the the Schr\"{o}dinger equation in generalized coordinates connected
with the QHJ equation through of ansatz: $\psi\left(\mathbf{q}\right)\equiv\exp\left(\frac{i}{\hbar}W\right).$
This expression is important because the zeros of the usual wave function $\psi\equiv\psi\left(q_{i},\bm{\kappa}\right)$
are the poles in the momentum function $p_{i}\left(q_{i},\bm{\kappa}\right)$.

Note that, we can explore the classical limit using the correspondence principle conveniently imposed, with $\hbar\rightarrow 0$, i.e.,
\[
\lim_{\hbar\rightarrow 0}\left[-\mathrm{i}\hbar\bm{\nabla}\cdot{\bf p}+{\bf p}\cdot{\bf p}\right]=2{\rm m}\left(E-V\left(\mathbf{q}\right)\right)\equiv{\bf p}^{c}\cdot{\bf p}^{c},
\]
where ${\bf p}^{c}$ is so called of classical momentum. 

\subsection{The Riccati's equation}

The CNLR equation is given by
\begin{equation}
\frac{d p_{i}}{d q}=P\left(q_i\right)+Q\left(q_i\right)p_{i}+R\left(q_i\right)p_{i}^{2},\label{eq:CRicatti}
\end{equation}
where $P\left(q_i\right)$, $Q\left(q_i\right)$, and $R\left(q_i\right)$ are complex value functions. The equation (\ref{eq:CRicatti}) always be reduced to a second order linear ordinary differential equation using
\begin{equation}
p_{i}=-\frac{1}{R\left(q_{i}\right)}\left(\frac{1}{u_{q_{i}}}\frac{d u_{q_{i}}}{d q_{i}}\right),\label{eq:psolution}
\end{equation}
where $R\left(q_i\right)$ is non-zero and differentiable \cite{Schuch-jphysconfseries504, Ince-book1956}. Note that, the equation (\ref{eq:psolution}) is fundamentally the relation (\ref{eq:ppsifunc}). In this form, we solve the following differential equation
\begin{equation}
\frac{d^2 u_{q_{i}}}{d q_{i}^2}-T\left(q_{i}\right)\frac{d u_{q_{i}}}{d q_{i}}+S\left(q_{i}\right)u_{q_{i}}=0,\label{eq:Schrodingereq}
\end{equation}
with 
\begin{equation}
S\left(q_{i}\right)=P\left(q_{i}\right)R\left(q_{i}\right),\quad T\left(q_{i}\right)=Q\left(q_{i}\right)+\frac{1}{R\left(q_{i}\right)}\left(\frac{d R\left(q_{i} \right)}{d q_{i}}\right).\label{eq:SandT}
\end{equation}
The equation (\ref{eq:Schrodingereq}) is the Schr\"{o}dinger equation, where $u_q$ are the eigenfunctions of the Hamiltonian. In this manner, $S\left(q_{i}\right)$ and $T \left(q_{i}\right)$ connect the solution of the Schr\"{o}dinger equation directly with the momentum function of QHJ equation through of relation (\ref{eq:psolution}).

\subsection{Action-angle variables}

At this point, we can define the quantum action variable \cite{Leacock-prl50, Leacock-pra33, Leacock-amjphys55, Bhalla-amjphys1997}, which is analog to the action variable in classical mechanics \cite{Goldstein-book2002} 
\begin{equation}
J_{q_{i}}=\frac{1}{2\pi}\oint_{\Gamma}dq_{i}\,p_{i},\label{eq:Jfunc}
\end{equation}
where $p_{i}\left(q_{i},\bm{\kappa}\right)\equiv p_i$ is a complex value function. Note that, from the equation (\ref{eq:psolution}), $p_{i}$ has a certain number of poles related with the nodes of the wave function. This fact implies that $J$'s are related directly with the zeros of eigenfunctions, $\psi$. Moreover, the integral (\ref{eq:Jfunc}) is closed in a $\Gamma$-contour in complex plane $\mathbb{C}$. $J_{q_{i}}$ is function of separation constants $\bm{\kappa}$ including the energy $E$\footnote{The energy can be degenerated or non-degenerated.}.  

The action variable is considerably studied in the literature \cite{Born-book1960, Leacock-amjphys55, Leacock-pra33, Pauling-book1985}, exploring deeply we can find a conjugated variable to $J_{q_{i}}$ by the Hellmann-Feynman Theorem for expected values \cite{Feynman-pr56}, i.e.,
\begin{equation}
w_{i}=\frac{\partial\left\langle \hat{K}\right\rangle }{\partial J_{q_i}}=\frac{\partial E }{\partial J_{q_i}},\label{eq:wangle}
\end{equation}
where $\hat{K}$ is the new Hamiltonian in terms of $J$'s, which is totally analog to classical mechanics. $w_{i}$ represent the frequencies of the system in classical mechanics, where the problem is reduced to know these frequencies in a periodic system \cite{Goldstein-book2002}. Note that, the $J$'s and $w$'s are functions and do not have the constraints of an operator.

In the sections below, we illustrate the traditional examples of quantum mechanics applying explicitly the formulation given above and show the advantages, and specially its simplicity of the computation.

\section{Harmonic oscillator}\label{sec:HarmonicOscillator}

\subsection{Solution of momentum function in terms of Hermite polynomials}
\begin{figure}[htbp]
	\centering
	\includegraphics[width=0.7\textwidth]{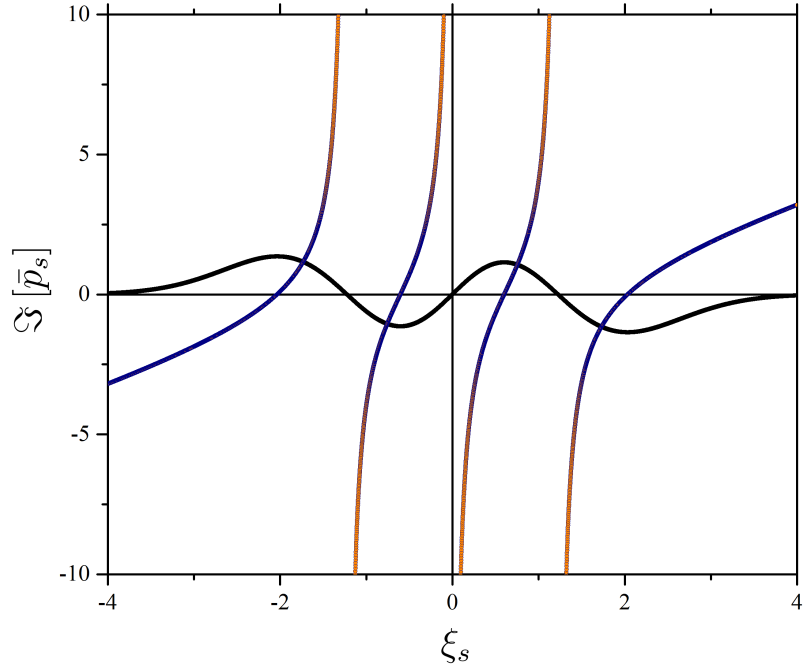}
	\caption{(Colour on-line) Imaginary part of momentum function $\bar{p}_s$ vs the conjugated coordinate $\xi_{s}$ for $n_{s}=3$ is shown in solid line with blue-orange gradient. The wave function $u_{s}$ is presented in black solid line in order to compare the position of nodes and anti-nodes with $\bar{p}_s$.}
	\label{fig::pxivsxi}
\end{figure}

The three-dimensional harmonic oscillator has the following potential
\[
V\left(\mathbf{q}\right)=\frac{1}{2}{\rm m}\left(\omega_{x}^{2}x^{2}+\omega_{y}^{2}y^{2}+\omega_{z}^{2}z^{2}\right).
\]
In this form, we obtain a set of differential equations from the equation (\ref{eq:HJEpfunction}) and they are denoted by
\begin{equation}
-\mathrm{i}\hbar\frac{dp_{s}}{ds}+p_{s}^{2}+m^{2}\omega_{s}^{2}s^{2}=\kappa_{s}, \label{eq:HJEHO}
\end{equation}
where $s$ represents each coordinate $x,y,z$ of $\mathbb{R}^{3}$. Observe that, the separation constants satisfy the condition $\kappa_{x}+\kappa_{y}+\kappa_{z}=2{\rm m}E$. In order to maintain the differential equation in an adimensional form, we conveniently substitute\footnote{From now on we use the notation $\bar{p}$ for the adimensional momentum function}
\begin{equation}
\xi_{s}=\sqrt{\lambda_{s}}s,\quad \bar{p}_{s}=\frac{p_{s}}{\sqrt{\lambda_{s}}\hbar},\label{eq:xipxi}
\end{equation} where $\lambda_{s}=\left(\hbar/{\rm m}\omega_{s}\right)^{1/2}$ is the natural length of the harmonic oscillator. Then, substituting (\ref{eq:xipxi}) into (\ref{eq:HJEHO}), we get
\begin{equation}
\frac{d\bar{p}_{s}}{d\xi_{s}}=\mathrm{i}\left(\frac{\kappa_{s}}{\lambda_{s}\hbar^{2}}-\xi_{s}^{2}\right)-\mathrm{i}\bar{p}_{s}^{2}.
\end{equation}
From the equation (\ref{eq:SandT}), we obtain 
\[
P\left(\xi_{s}\right)=\mathrm{i}\left(\frac{\kappa_{s}}{\lambda_{s} \hbar^{2}}-\xi_{s}^{2}\right),\quad Q\left(\xi_{s}\right)=0,\quad R\left(\xi_{s}\right)=-\mathrm{i},
\]
\[
S\left(\xi_{s}\right)=\left(\frac{\kappa_{s}}{\lambda_{s} \hbar^{2}}-\xi_{s}^{2}\right),\quad T\left(\xi_{s}\right)=0.
\]
Thus, one finds the equation (\ref{eq:Schrodingereq})\cite{Fluegge-book1994} 
\begin{equation}
\frac{d^{2}u_{s}}{d\xi_{s}^{2}}+\left(\frac{\kappa_{s}}{\lambda_{s} \hbar^{2}}-\xi_{s}^{2}\right)u_{s}=0,\label{eq:SeqHO}
\end{equation}
where the solution is 
\begin{equation}
u_{s}=A H_{n_{s}}\left(\xi_{s}\right){\rm e}^{-\frac{1}{2}\xi_{s}^{2}},\label{eq:uxi}
\end{equation}
for $\kappa_{s}/\hbar^{2}\lambda_{s}=2n_{s}+1$, with $n_{s}\in\mathbb{Z}^{+}\cup\left\{ 0\right\} $. $A$ is the normalization constant that depends on $n_{s}$ and $H_{n_{s}}\left(\xi_{s}\right)$ are the Hermite polynomials. Note that, $u_{s}$ is asymptotically well-behaved. Following the sequence from (\ref{eq:psolution}) is easy to see that the momentum function is: 
\begin{equation}
\bar{p}_{s}=-\mathrm{i}\left[\xi_{s}-\frac{H_{n_{s}+1}\left(\xi_{s}\right)}{H_{n_{s}}\left(\xi_{s}\right)}\right],\label{eq:ps}
\end{equation}
The equation (\ref{eq:ps}) is the ratio between two consecutive Hermite polynomials. An illustration of this momentum function is shown in Figure \ref{fig::pxivsxi} for $n_s=3$.

\subsection{Nodes of harmonic oscillator}
The $p_{s}$ solution is intuitively correct, in the Figure \ref{fig::pxivsxi} we can observe the behavior of the momentum function, we see that can be infinite or cross the zero on the abscissa. Comparing for the same quantum number, we can see that correspond precisely with the zeros and maximums of the wave function. In this way, we consider the solution (\ref{eq:ps}) and define the action variable as
\begin{equation}
J_{s}=\frac{1}{2\pi}\oint ds \,p_{s} =\frac{\hbar}{2\pi}\oint d\xi_{s}\,\bar{p}_{s},\label{eq:Js} 
\end{equation}
Note that the contour integral is defined on the disk of radius $R>0$, which contains the roots of Hermite polynomials $H_{n_{s}}$ \cite{Greenwood-bullamermathsoc54}. From the right hand side of the equation (\ref{eq:Js}) we have
\[
J_{s}=-\frac{\mathrm{i}}{2\pi}\oint d\xi_{s}\,\xi_{s}+\frac{\mathrm{i}}{2\pi}\oint d\xi_{s}\,\frac{H_{n_{s}+1}\left(\xi_{s}\right)}{H_{n_{s}}\left(\xi_{s}\right)},
\]
and applying the residues Theorem, we then have that the first integral is null and the second one is 
\[
\oint d\xi_{s}\,\frac{H_{n_{s}+1}\left(\xi_{s}\right)}{H_{n_{s}}\left(\xi_{s}\right)}=-2\pi \mathrm{i} n_{s}.
\]
At last, we conclude that
\begin{equation}
J_{s}=n_{s}\hbar.\label{eq:Jscounter}
\end{equation}
Observe that $n_s\in \mathbb{Z}^{+}\cup \left\{0\right\}$ are number of nodes of the Hermite polynomials $H_{n_s}\left(\xi_{s}\right)$ and therefore are the number of nodes of wave function multiplied by $\hbar$. Moreover, we can obtain exactly the anti-nodes for the harmonic oscillator integrating the inverse of $p_{s}$ over a closed trajectory on the complex plane $\mathbb{C}$.

We can easily show that the energy is a function of variable $J_{s}$, thus
\begin{equation}
E=\sum_{s}\omega_{s}\left(J_{s}+\frac{\hbar}{2}\right).\label{eq:EHO}
\end{equation}
Using the Hellmann\textendash Feynman Theorem, the expected value of connect the action-variable with the angle variable, i.e.,
\[
w_{s}=\omega_{s},
\]
It implies that has the same form of the classical variable \cite{Goldstein-book2002}. This fact corresponds to another natural result of the solution, i.e., the physical meaning of $w_{s}$ is the natural frequency of the system.

\section{Hydrogen atom}\label{sec:HydrogenAtom}

\subsection{Solution of momentum functions for Coulomb potential}
The next system to be studied and not least important is the Coulomb potential, which is a degenerate problem by definition, in both quantum and classical mechanics. We consider the following usual potential
\[
V\left(\mathbf{q}\right)=-\frac{k}{r},
\]
where $k=e_{1}e_{2}$ with $e_{1},e_{2}>0$. Following the notation given in (\ref{eq:HJEpfunction}) we have the set of equations
\begin{equation}
-\mathrm{i}\hbar\frac{dp_{\phi}}{d\phi}+p_{\phi}^{2}=\kappa_{\phi},\label{eq:HJepphi}
\end{equation}
\begin{equation}
-\frac{\mathrm{i}\hbar}{\sin\theta}\frac{d}{d\theta}\left(\sin\theta p_{\theta}\right)+p_{\theta}^{2}=\kappa_{\theta}-\frac{\kappa_{\phi}}{\sin^{2}\theta},\label{eq:HJeptheta}
\end{equation}
\begin{equation}
-\frac{\mathrm{i}\hbar}{r^{2}}\frac{d}{dr}\left(r^{2}p_{r}\right)+p_{r}^{2}=2 \mathrm{m} \left(E+\frac{k}{r}\right)-\frac{\kappa_{\theta}}{r^{2}}.\label{eq:HJpr}
\end{equation}
where $\kappa_\theta$, $\kappa_\phi$ and $E$ are the separation constants.

\subsubsection{Solution corresponding to $\phi$}

We rewrite the equation (\ref{eq:HJepphi}) in an adimensional form:
\begin{equation}
\frac{d\bar{p}_{\phi}}{d\phi}=\frac{\mathrm{i} \kappa_{\phi}}{\hbar^{2}}-\mathrm{i}\bar{p}_{\phi}^{2},
\end{equation}
where
\begin{equation}
\bar{p}_\phi=\frac{p_{\phi}}{\hbar}\label{eq:phipphi}
\end{equation}
Note that the function $p_{\phi}$ has units of angular momentum. From the functions established in  (\ref{eq:SandT}), we have
\[
P\left(\phi\right)=\frac{\mathrm{i} \kappa_{\phi}}{\hbar^{2}},\quad Q\left(\phi\right)=0,\quad R\left(\phi\right)=-\mathrm{i}
\]
\[
S\left(\phi\right)=\frac{\kappa_{\phi}}{\hbar^{2}},\quad T\left(\phi\right)=0.
\] 
Substituting the last expressions into (\ref{eq:Schrodingereq}), we obtain
\begin{equation}
\frac{d^{2}u_{\phi}}{d\phi^{2}}+\frac{\kappa_{\phi}}{\hbar^{2}}u_{\phi}=0,
\end{equation}
whose solution is well known: 
\begin{equation}
u_{\phi}=B {\rm e}^{\mathrm{i} m\phi}\label{eq:uphi}
\end{equation}
for $\kappa_{\phi}^{2}/\hbar^{2}  =m^{2}$, with $m\in\mathbb{Z}$. $B$ is a normalization constant. Therefore, the momentum function
\begin{equation}
\bar{p}_{\phi}=m.\label{eq:pphi}
\end{equation}
As expected $p_{\phi}$ is a constant of motion; in fact, we can see that in the expression (\ref{eq:HJepphi}) $\phi$ does not appear explicitly.

\subsubsection{Solution corresponding to $\theta$}

\begin{figure}[htbp]
	\centering
	\includegraphics[width=0.7\textwidth]{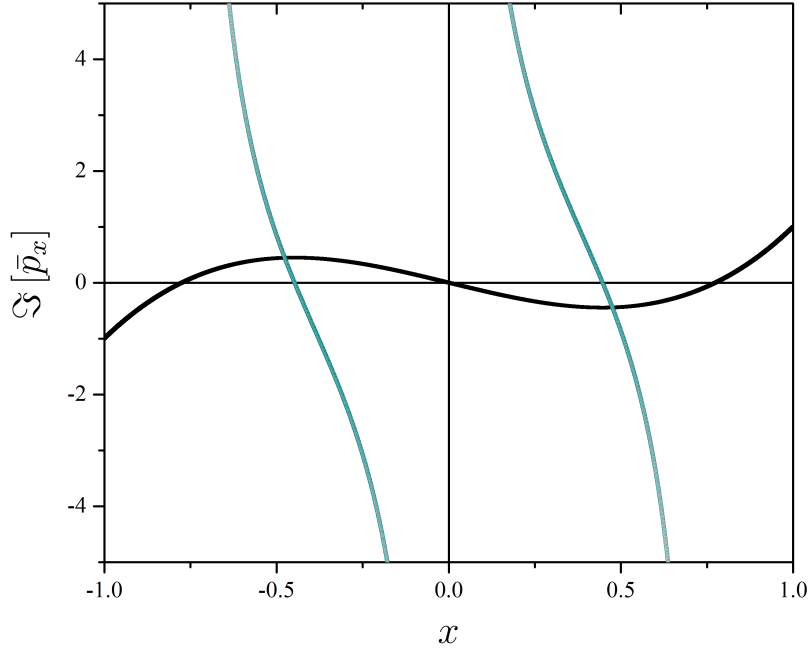}
	\caption{(Colour on-line) The solid line with green-gray gradient shows the imaginary part of momentum function $\bar{p}_x$ vs the conjugated coordinate $x$ for $\ell=3$ and $m=0$. The wave function $u_{x}$ is presented in black solid line in order to compare the position of nodes and anti-nodes with $\bar{p}_x$.}
	\label{fig::pxvsx}
\end{figure}

From the equation (\ref{eq:HJeptheta}) we choose a suitable adimensional form given as
\begin{equation}
\frac{d\bar{p}_{x}}{dx}=-\frac{\mathrm{i}}{1-x^{2}}\left(\frac{\kappa_{\theta}}{\hbar^2}-\frac{m^{2}}{1-x^{2}}\right)+\frac{2x}{1-x^{2}}\bar{p}_{x}+\mathrm{i}\bar{p}_{x}^{2}, \label{eq:adpx}
\end{equation}
where
\begin{equation}
x=\cos\theta,\quad \bar{p}_x=\frac{p_{\theta}}{\hbar \sin \theta}.\label{eq:xpx}
\end{equation}
We recognize that
\[
P\left(x\right)=-\frac{\mathrm{i}}{1-x^{2}}\left(\frac{\kappa_{\theta}}{\hbar^2}-\frac{m^{2}}{1-x^{2}}\right), \quad Q\left(x\right)=\frac{2x}{1-x^{2}},\quad R\left(x\right)=\mathrm{i}.
\]
Observe that $R\left(x \right)$ is non-zero and differentiable for all $x\in\left(-1,1\right)$. By the equation (\ref{eq:SandT}) we have the functions
\[
S\left(x\right)=\frac{1}{1-x^{2}}\left(\frac{\kappa_{\theta}}{\hbar^2}-\frac{m^{2}}{1-x^{2}}\right),\quad T\left(x\right)=\frac{2x}{1-x^{2}}.
\] 
Then, the equation (\ref{eq:adpx}) becomes
\begin{equation}
\left(1-x^{2}\right)\frac{d^{2}u{}_{x}}{dx^{2}}-2x\frac{du{}_{x}}{dx}+\left(\frac{\kappa_{\theta}}{\hbar^2}-\frac{m^{2}}{1-x^{2}}\right)u_{x}=0
\end{equation}
where $\kappa_{\theta}/\hbar^2=\ell\left(\ell+1\right)$, with $\ell=\left|m\right|,\left|m\right|+1,\dots.$ The solution of the differential equation is given
by 
\begin{equation}
u_{x}= C P_{\ell}^{\left|m\right|}\left(x\right),\label{eq:ux}
\end{equation}
where $C$ is a normalization constant depending on $\ell$ and $\left|m\right|$. The functions $P_{\ell}^{\left|m\right|}\left(x\right)$ are the associated Legendre functions \cite{Erdelyi-book1955v1, Erdelyi-book1955v2, Bauer-mathcomp46}. From the relation (\ref{eq:psolution}) we obtain the solution
\begin{equation}
\bar{p}_{x}=\mathrm{i} \left[-\frac{\ell x}{1-x^{2}}+\frac{\left(\ell+\left|m\right|\right)}{1-x^{2}}\frac{P_{\ell-1}^{\left|m\right|}\left(x\right)}{P_{\ell}^{\left|m\right|}\left(x\right)}\right],\label{eq:px}
\end{equation}
and using the simple recurrence relation for its derivative:
\[
\left(1-x^{2}\right)\frac{d P_{\ell}^{\left|m\right|}\left(x\right)}{dx}=-\ell xP_{\ell}^{\left|m\right|}\left(x\right)+\left(\ell+\left|m\right|\right)P_{\ell-1}^{\left|m\right|}\left(x\right).
\]
Figure \ref{fig::pxvsx} shows an example of the momentum function for $\ell=3$ and $m=0$.

\subsubsection{Solution corresponding to $r$}

\begin{figure}[htbp]
	\centering
	\includegraphics[width=0.7\textwidth]{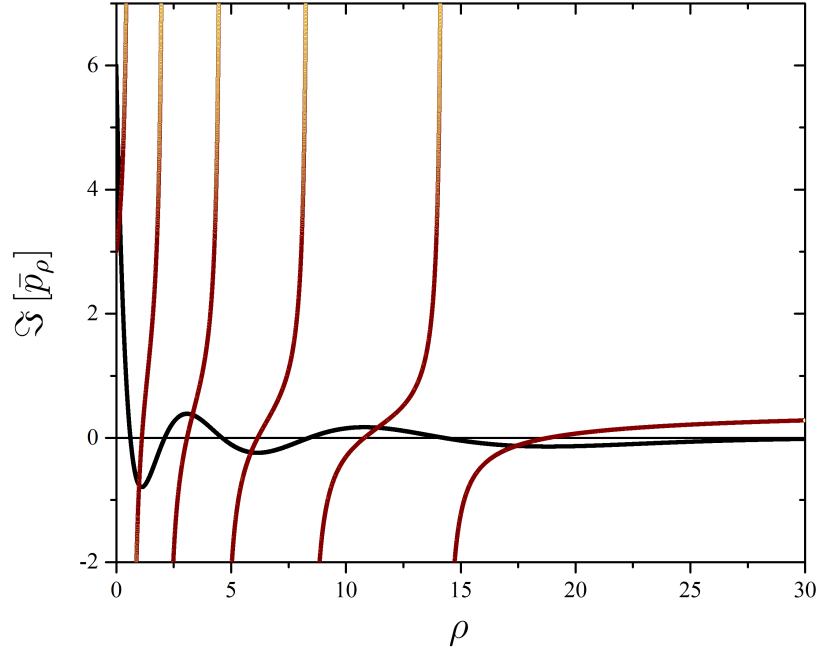}
	\caption{(Colour on-line) The solid line with red-yellow gradient shows imaginary part of momentum function $\bar{p}_{\rho}$ vs the conjugated coordinate $\rho$ for $n=6$ and $\ell=0$. The wave function $u_{\rho}$ is presented in black solid line in order to compare the position of nodes and anti-nodes with $\bar{p}_{\rho}$.}
	\label{fig::prhovsrho}
\end{figure}

Finally, taking (\ref{eq:HJpr}) we get
\begin{equation}
\frac{d\bar{p}_{\rho}}{d\rho}=\mathrm{i}\left(-\frac{1}{4}+\frac{\lambda}{\rho}-\frac{\kappa_{\theta}}{\hbar^2 \rho^{2}}\right)-\frac{2}{\rho}\bar{p}_{\rho}-\mathrm{i}\bar{p}_{\rho}^{2},
\end{equation}
where
\begin{equation}
\rho=2\alpha r, \quad \bar{p}_{\rho}=\frac{p_{r}}{2\alpha\hbar},\label{eq:rhoprho}
\end{equation}
with $\alpha^{2}=-2{\rm m}E/\hbar^{2}$ and $\lambda={\rm m}k/\alpha\hbar^{2}$. Note that, the function $R\left(\rho \right)$ is non-zero and differentiable for all $\rho \in\left(0,+\infty\right)$, and $\kappa_{\theta}/\hbar^2=\ell\left(\ell+1\right)$. Thus,
\[
S\left(\rho\right)=\left(-\frac{1}{4}+\frac{\lambda}{\rho}-\frac{\ell\left(\ell+1\right)}{\rho^{2}}\right),\quad T\left(\rho\right)=-\frac{2}{\rho}.
\]
Then we obtain the following differential equation
\begin{equation}
\frac{d^{2}u{}_{\rho}}{d\rho^{2}}+\frac{2}{\rho}\frac{du{}_{\rho}}{d\rho}+\left(-\frac{1}{4}+\frac{\lambda}{\rho}-\frac{\ell\left(\ell+1\right)}{\rho^{2}}\right)u_{\rho}=0,
\end{equation}
and its solution is
\begin{equation}
u_{\rho}=D \mathrm{e}^{-\rho/2}\rho^{\ell}L_{n-\ell-1}^{2\ell+1}\left(\rho\right),\label{eq:urho}
\end{equation}
where $D$ is a constant depending on $\ell$ and $n$, with $n \in \mathbb{Z}$. The functions $L_{n+\ell}^{2\ell+1}\left(\rho\right)$ are the associated Laguerre polynomials \cite{Aizenshtadt-book1966}. Following the expression (\ref{eq:psolution}) we have
\begin{equation}
\bar{p}_{\rho}=\mathrm{i}\left[\frac{1}{2}-\frac{\ell}{\rho}+\frac{L_{n-\ell-2}^{2\ell+2}\left(\rho\right)}{L_{n-\ell-1}^{2\ell+1}\left(\rho\right)}\right].\label{eq:prho}
\end{equation}
The Figure \ref{fig::prhovsrho} illustrates $\bar{p}_{\rho}$ for $n=6$ and $\ell=0$. In contrast $p_{\phi}$ and $p_{\theta}$, we can see that the solution of $p_r$ has units of linear momentum.

\subsection{Nodes of hydrogen atom}

According on that line of thought, we propose the action variables for the Coulomb potential. We start with the simplest form, which obviously corresponds to the constant of motion $p_{\phi}$, i.e.,
\begin{equation}
J_{\phi}=\frac{1}{2\pi}\oint d\phi\,p_{\phi}=\frac{\hbar}{2 \pi}\oint d\phi \, \bar{p}_{\phi}
\end{equation}
Substituting (\ref{eq:pphi}) into $J_{\phi}$, we get
\begin{equation}
J_{\phi}= m \hbar.\label{eq:Jphicounter}
\end{equation}
where $m$ besides being the magnetic quantum number also represents the number of angular nodes around the $\phi$-axis.

Remember that the expression (\ref{eq:px}) for action variable in $\theta$ is given by
\begin{equation}
J_{\theta}=\frac{1}{2\pi}\oint d\theta\,p_{\theta}=-\frac{\hbar}{2 \pi}\oint dx \, \bar{p}_{x}.
\end{equation}
So,
\[ 
J_{\theta}=-\frac{\mathrm{i} \ell}{2 \pi}\oint dx \frac{x}{1-x^{2}}+\frac{\mathrm{i}\left(\ell+\left|m\right|\right)}{2\pi} \oint dx \frac{1}{1-x^{2}}\frac{P_{\ell-1}^{\left|m\right|}\left(x\right)}{P_{\ell}^{\left|m\right|}\left(x\right)}.
\]
Resolving by residues Theorem, we can show the first integral is $-2\pi \mathrm{i}$, and another one is
\[
\oint dx \frac{1}{1-x^{2}}\frac{P_{\ell-1}^{\left|m\right|}\left(x\right)}{P_{\ell}^{\left|m\right|}\left(x\right)}=0.
\]
for all $\ell$ and $m$. Therefore,
\begin{equation}
J_{\theta}=\ell\hbar.\label{eq:Jthetacounter}
\end{equation}
$\ell$ is the angular momentum quantum number or number of angular nodes.

At last, the integral for (\ref{eq:prho}) is
\begin{equation}
J_{r}=\frac{1}{2 \pi}\oint dr \,p_{r}= \frac{\hbar}{2 \pi}\oint d\rho\,\bar{p}_{\rho}.
\end{equation}
For this case,
\[
J_r=\frac{\mathrm{i}}{4\pi} \oint d\rho -\frac{\mathrm{i \ell}}{2\pi} \oint \frac{d\rho}{\rho}+\frac{\mathrm{i}}{2\pi} \oint d\rho\frac{L_{n-\ell-2}^{2\ell+2}\left(\rho\right)}{L_{n-\ell-1}^{2\ell+1}\left(\rho\right)}.
\]
Solving each integral, we obtain the first one is null, the second one is $2 \pi \mathrm{i}$, and the last one is
\[
\oint d\rho\frac{L_{n-\ell-2}^{2\ell+2}\left(\rho\right)}{L_{n-\ell-1}^{2\ell+1}\left(\rho\right)}=-2\pi \mathrm{i}\left(n-\ell-1\right),
\]
for all $n$ and $\ell$ specified for each wave function. In this manner, we get
\begin{equation}
J_{r}=\left(n-1\right)\hbar.\label{eq:Jrhocounter}
\end{equation}
This result shows the number of total nodes of wave function of Hydrogen atom, in terms of principal quantum number. As expected the number of nodes is zero in the ground state.

Finally, from the solution $\lambda=n$ we can compute the energy as function of $J_{r}$ as
\begin{equation}
E=-\frac{{\rm m}k^{2}}{2}\frac{1}{\left(J_{r}+\hbar\right)^{2}}.\label{eq:ECoulomb}
\end{equation}
The energy derived by Bohr for each orbit of the hydrogen atom. On the other hand, we have
\[
w_{r}=\frac{{\rm m}k^{2}}{\left(J_{r}+\hbar\right)^{3}}
\]
is the angle variable. Note that, we obtain the classical limit from the last equations doing $\hbar\rightarrow 0$ \cite{Goldstein-book2002}.

\section{Conclusions}\label{sec:conclusion}

In the present article, we have shown the exact solution of QHJ equation, investigating the most important bound state cases for $3$D problem. We observe that the dynamics of generalized canonical momentum functions in the HJ formulation are completely analog to dynamics of the wave functions in the Schr\"{o}dinger equation. Besides, we establish a clear connection and equivalence between the QHJ equation and Schr\"{o}dinger equation using CNLR equation. Moreover, we show that the goodness of HJ not tackle the problem of eigenvalues without solving the equations of motion, but avoid the problem of operators using generalized functions containing embedded dynamics of the system. 

In addition, the exploration the angle-action variables through the explicit solution of the equation of motion. For a bound state problem we demonstrate that where $p_{i}$ is non-analytic we have $0,1,2...$ nodes which corresponds to ground state, first excited state, second excited state, etc. In other words, from the point of view of quantum mechanics the $J$'s are \textit{counters of nodes} of the wave function. Moreover, we can obtain exactly the anti-nodes for integrating the inverse of momentum function over a closed trajectory on the complex plane $\mathbb{C}$. On the other hand, the method to compute the action variable open up interesting mathematical properties such the ratio between two consecutive polynomials.

\section*{Acknowledgments}
L.A. Poveda-Cuevas would like to thank Instituto de Matem\'{a}tica e Estat\'{i}stica, Universidade de S\~{a}o Paulo and CAPES (Brazil) for the PhD Fellowship support. F.J. Poveda-Cuevas thanks SECITI-CLAF -- SECITI 064/2015 -- and DGAPA (M\'{e}xico) for the postdoctoral fellowship.

\section*{References}
\bibliographystyle{iopart-num}
\bibliography{bibqhj}

\end{document}